\documentclass[amsmath,amssymb,prl,twocolumn,superscriptaddress]{revtex4}

\usepackage[utf8]{inputenc}
\usepackage{amsmath}
\usepackage{graphicx}
\usepackage{chemformula}
\usepackage{hyperref}
\usepackage{enumerate}
\usepackage{xcolor}
\usepackage{enumitem}
\newcommand{\ciceco}{CICECO - Instituto de Materiais de Aveiro, Department of Chemistry, 3810-193 Aveiro, Portugal}

\begin{document}
\title{Topological edge states in a double isomeric Class-II oligo(indenoindene)}

\author{Ricardo Ortiz}\thanks{Corresponding author: ricardo.ortiz.cano@ua.pt}
\affiliation{\ciceco}

\begin{abstract}
I report the theoretical prediction of non-trivial physics in a one dimensional multiradical system consisting in fused six and five membered $\pi$-conjugated carbon rings, known as oligo(indenoindene) (OInIn). Topologically protected electronic states may emerge in fermionic chains if there is an alternation in the coupling of adjacent unpaired electrons, being described effectively by the Su-Schrieffer-Heeger (SSH) model. Class-II OInIn isomers act as tight-binding chains in the non-interacting regime, thus we can expect the emergence of SSH physics in an OInIn produced by the combination of two isomers that belong to this class. That is the case of the system studied in this manuscript, whose calculated non-interacting band structure shows a gap opening compared to the gapless pure isomeric forms, hosting ingap localized states at the chain termini depending on the termination, and a non-zero Zak phase that confirms the non-trivial topology. These results were consistent with spin unrestricted mean-field Hubbard and density functional theory calculations, showing antiferromagnetic unquenched local magnetic moments at the pentagons, and strong edge localization depending on the termination. This work advances in the understanding of the physics of non-alternant multiradical $\pi$-conjugated hydrocarbons.    
\end{abstract}
\maketitle

\section{Introduction}

The search for new platforms hosting topologically protected boundary states is one of the main quests in condensed matter physics and materials science\cite{qi2011topological,senthil2015symmetry,verresen2017one}.
In this sense, graphene nanoribbons have been established as versatile platforms to realize the low energy physics of model 1D Hamiltonians\cite{massivefermions2020,crommiescience,groning2018engineering}, where the rational design of their edges promotes the emergence of strongly localized modes, ranging from flat bands localized at zigzag edges\cite{nakada1996edge,yazyev2011theory,Ruffieux16,deniz2025electronic} to localized edge states in finite graphene nanoribbons\cite{wang2016,rizzo2018topological,li2021topological}. 

The implementation of the on-surface synthesis technique, in combination with ultra-high vacuum conditions and STM spectroscopy, has permitted the exploration of such localized states in graphene nanoribbons\cite{talirz2016surface,houtsma2021atomically,ORTIZ2020100595}. By engineering their width and shape, it is possible to design unit cells that belong to different topological classes, 
as it was derived by S. Louie et al. in their seminal work\cite{louiejunction}. As a consequence, junctions that alternate fragments with different Berry phase ($\gamma$) will host a localized state at the interface, which opens up the possibility of designing 1D periodic systems described by the SSH model\cite{su1979solitons}:

\begin{equation}
{\cal H}_{SSH} = \sum_{m}(t_1 a^\dagger_{m}b_{m} + t_2 a^\dagger_{m+1}b_{m} + h.c.),
\label{eq1}
\end{equation}     

where $a^\dagger_{m}$ ($a_{m}$) and $b^\dagger_{m}$ ($b_{m}$) are the creation (annihilation) operators in second quantization for two orbitals at the $m^{th}$ unit cell.

Thus, by the modulation of the $t_1/t_2$ ratio in Eq.\ref{eq1}, two different topological regimes are connected by a gap closure at $t_1=t_2$. Both of these situations ($t_1>t_2$ or $t_1<t_2$) can be well characterized by the value of $\gamma$, which receives the name of Zak-phase (${\cal Z}$) in 1D systems\cite{zak1989berry}, and can only take values of $0$ or $\pi$ (modulo $2\pi$) if the system presents inversion/mirror symmetry. This $\mathbb{Z}_2$ topological invariant can be directly calculated for each band in the 1D band structure:

\begin{equation}
{\cal Z}_n =  \int_{\bold{BZ}} i\langle \varphi_n({k}) | \partial_{k} \varphi_n({k}) \rangle d{k},
\end{equation}  

where $n$ labels the band and ${k}$ is the wavevector.
As a consequence, graphene  nanoribbons that belong to the topologically non-trivial regime (${\cal Z}=\pi$) are expected to host ingap edge states, as it has been probed with STM in several cases described in the bibliography\cite{massivefermions2020,crommiescience,groning2018engineering}.

Another possibility of engineering unpaired electrons in graphene nanostructures is by the inclusion of pentagonal defects in their backbone\cite{hu2016toward}, where a minimalistic example consists in a ladder polymer that alternates hexagons and pentagons known as oligo(indenoindenes)\cite{transchainexp} (OInIn). Such one dimensional platform has six isomers, that can be divided into two different classes depending on the dominant magnetic interactions\cite{ortiz2023magnetic}:

\begin{enumerate}[label=(\Roman*)]
        \item Competing ferromagnetic and antiferromagnetic first-neighbors exchange and antiferromagnetic second-neighbors exchange.
        \item Antiferromagnetic first-neighbors exchange.
\end{enumerate}  

\begin{figure*}
 \centering
    \includegraphics[width=0.9\textwidth]{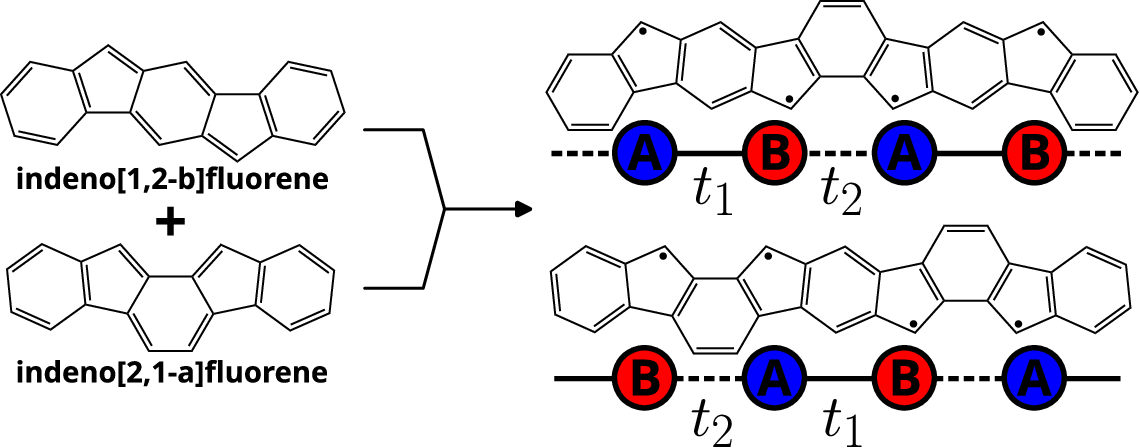}
\caption{Chemical structure of the systems of interest: (left) the pure isomeric forms of two Class-II indenofluorene isomers, which combine into (right) a double isomeric Class-II oligo(indenoindene).}
%\label{Structure}
\label{fig1}
\end{figure*}

Some pure isomeric forms of OInIn have been synthesized on surfaces in the recent years\cite{mishra2024bistability,truxenewang,truxene,transchainexp,villalobos2025globally}, as well as one dimensional chains of indenofluorenes\cite{intercalationsJACS,di2022surface}, and solution-synthesized derivatives\cite{dressler2017synthesis,shimizu2013indeno,nishida2012synthesis,martinez2024configurationally,shimizu2011indeno}. However, to the best of my knowledge, neither an experimental characterization or a theoretical description of the physics of ladder polymers consisting in the combination of two or more isomers has been reported. In this work, I present a series of calculations that model the low energy electronic and magnetic properties of a one dimensional system conformed by the repetition of two Class-II OInIn (Fig.\ref{fig1}), predicting the presence of topologically protected states at the edges of finite fragments.

\section{Computational details}

In this work, a series of calculations are done at different levels of theory. First, I model the non-interacting regime with a simple tight-binding model:

\begin{equation}
{\cal H}_t = t\sum_{\langle i,j\rangle} (c^\dagger_ic_j + h.c.),
\end{equation}

where $c^\dagger_i$($c_i$) creates(annihilates) an electron at orbital $i$, and coupling is counted by the hopping parameter ($t=-2.7$eV) for each pair of first-neighouring orbitals $\langle i,j\rangle$. I assume that the molecule is entirely planar, thus $\pi$ and $\sigma$ orbitals are orthogonal and we can approximate the model to just consider one $\pi$-orbital per carbon atom. Then, the 1D band structures are calculated deriving a tight-binding crystal Hamiltonian using the Bloch's theorem ${\cal H} (k) = {\cal H}_t + V_t e^{ik|\bold{R}|} + V^T_t e^{-ik|\bold{R}|}$, where $|\bold{R}|\equiv R_x$ is the modulus of the 1D crystal vector, and $V_t$ is the intercell hopping matrix. 

Second, electronic interactions are included in the form of an on-site Coulomb repulsion by the Hubbard model, that is replaced by the interaction between electrons and an effective background potential self-consistently calculated in a mean-field approximation:

\begin{equation}
{\cal H}_{MF} = U\sum_i (n_{i\uparrow}\langle n_{i\downarrow}\rangle + n_{i\downarrow}\langle n_{i\uparrow}\rangle),
\end{equation}

where $U$ is the Hubbard parameter that accounts for the Coulombic repulsion, $n_{i\sigma}=c^\dagger_{i\sigma}c_{i\sigma}$ counts electrons at orbital $i$, $\langle n_{i\sigma}\rangle=\sum_{n}f_{n\sigma}|\varphi_{n\sigma}(i)|^2$ is the electron density with spin $\sigma\in \{\uparrow,\downarrow\}$ and $f_{n\sigma}$ is a filling factor. Interestingly, the mean-field Hubbard model permits the convergence of solutions with spin broken symmetry\cite{Yazyev_2010,LADO201556,Carbonbasednanostructures}, where the magnetization in spin units is just half the spin density:

\begin{equation}
m_i = \frac{\langle n_{i\uparrow}\rangle-\langle n_{i\downarrow}\rangle}{2},
\end{equation}

and the total energy is given by:

\begin{equation}
E_{MF} = \sum_{n,\sigma} (f_{n\sigma}\varepsilon_{n\sigma}) - U \sum_i (\langle n_{i\uparrow}\rangle\langle n_{i\downarrow}\rangle).
\end{equation}

Third, I obtained optimized geometries of finite molecules using density functional theory (DFT) with the ORCA 5.0.4 package\cite{orcageneric,neese2022software}. The results showed no significant deviation from planarity. Moreover, vibrational frequency calculations for the molecules with $P=2,\ 3$ and $4$ pentagons confirmed that their geometries correspond to a local minimum. I selected the lowest spin multiplicities, i.e. $S=0$ ($S=1/2$) for OInIn with even (odd) number of pentagons. I used the PBE0 density functional\cite{adamo1999toward} with the RIJCOSX approximation\cite{neese2009efficient}, including the def2-SVP basis\cite{weigend2005balanced}, the def2/J auxiliary basis\cite{weigend2006accurate}, and the D3BJ method for dispersion interactions\cite{grimme2011effect}. The relaxed geometries and unrestricted Kohn-Sham orbitals were used as inputs for a CASSCF calculation\cite{roos1980complete}, using the def2/C auxiliary basis\cite{hellweg2007optimized}, and converging Active Spaces with a smallest fractional occupation with no more than $0.09$ electrons and the highest with no less than $1.91$ electrons.

\section{Results and discussion}

In graphene nanostructures, unpaired electrons manifest as states pinned at the Fermi energy\cite{ortiz19} ($E_F$) in the non-interacting spectrum. When two of these states ($\varphi_z(i)$) are connected by hopping\cite{Ortiz18}, hybridization causes an energy splitting that is equal to twice the hopping matrix element:

\begin{equation}
\delta = 2\langle \varphi_z(i) | {\cal H}_t | \varphi_{z'}(i)\rangle.
\end{equation}

\begin{figure}
 \centering
    \includegraphics[width=0.48\textwidth]{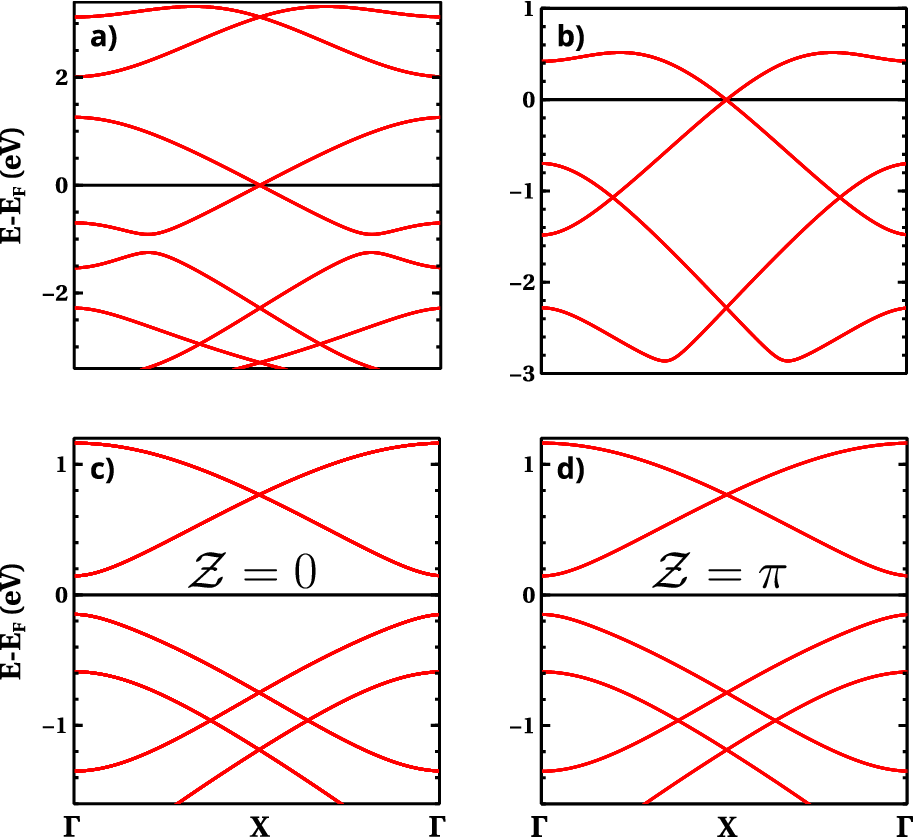}
\caption{Band structure of the OInIn polymers, computed with a first-neighbours tight-binding model ($t=-2.7$eV), and derived from a) indeno[1,2-b]fluorene, b) indeno[2,1-a]fluorene and their combinations whose finite fragments c) does not host and d) host edge states.}
%\label{Structure}
\label{fig2}
\end{figure}

\begin{figure}[h!]
 \centering
    \includegraphics[width=0.48\textwidth]{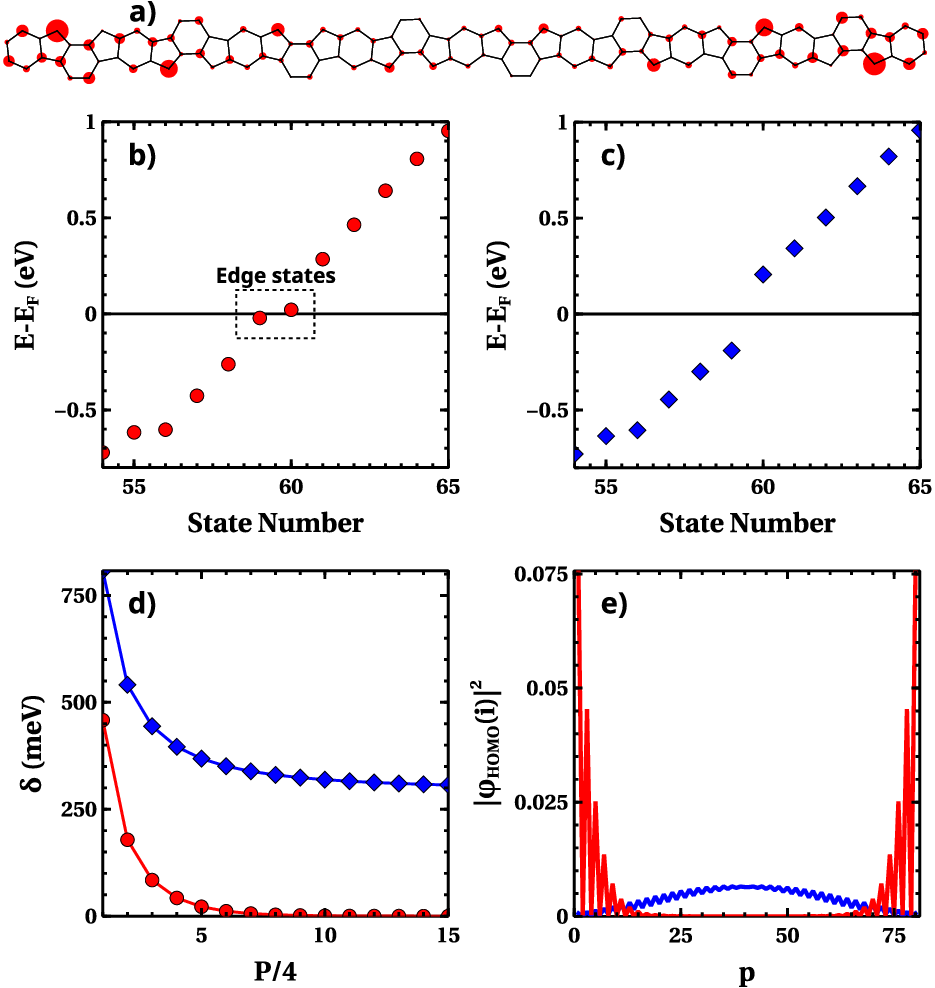}
\caption{a) Representation of $|\phi_{HOMO}(i)|^2$ in an OInIn with a termination that hosts edge states and b) its tight-binding espectrum. c) Tight-binding spectrum for an OInIn with the same number of pentagons as in the previous panels, but with a termination with no edge states. d) HOMO-LUMO gap ($\delta$) evolution with the number of pentagons ($P$). e) Distribution of $|\phi_{HOMO}(i)|^2$ at each pentagon vertex $p$ for an OInIn with each termination and $P=80$. Red (blue) are calculations for a ribbon with (without) edge states.}
%\label{Structure}
\label{fig3}
\end{figure}

If we consider a system with $P$ hybridized unpaired electrons, non-interacting physics would be described by a one dimensional tight-binding chain governed by an effective hopping ($\tilde{t}=\delta/2$). In Figure\ref{fig2} we demonstrate the former hypothesis by calculating the tight-binding band structure for Class-II OInIn isomers, where $P$ coincides with the number of pentagons. The first two panels (Fig.\ref{fig2}a,b) correspond to periodic OInIn extended from the pure isomers shown in Fig.\ref{fig1}, where the effective coupling between adjacent $\varphi_z(i)$ yields two crossing bands with linear band dispersion close to $E_F$\cite{ortiz2023magnetic}. This is also the case of the SSH model for $t_1=t_2$, which is nothing else than the one-dimensional tight-binding model with two atoms in the unit cell, depicting a Dirac cone at $X\equiv k=\pi/R_x$.

Modulation of $t_1$ and $t_2$  opens up a gap ($t_1\neq t_2$) in the otherwise gapless band structure of the SSH model. In spite the eigenvalues are the same as long as $ t_1/t_2 = t'_2/t'_1$, the wave functions are different in both hopping regimes ($t_1 > t_2$ and $t'_1<t'_2$), representing trivial and non-trivial topological phases with ${\cal Z} = 0$ and ${\cal Z} = \pi$, respectively. Following this, the effective hopping between two unpaired pentagon electrons is going to depend on their relative orientation, thus a polymer consisting in a combination of two different OInIn Class-II isomers will potentially host the same physics as the SSH model. In the case of the indeno[2,1-a]fluorene pentagons orientation the effective hopping ($\tilde{t}$) is smaller than for the indeno[1,2-b]fluorene orientation, as can be inferred from the calculated $\delta$ in tight-binding $P=2$ graphs.

\begin{figure*}
 \centering
    \includegraphics[width=1\textwidth]{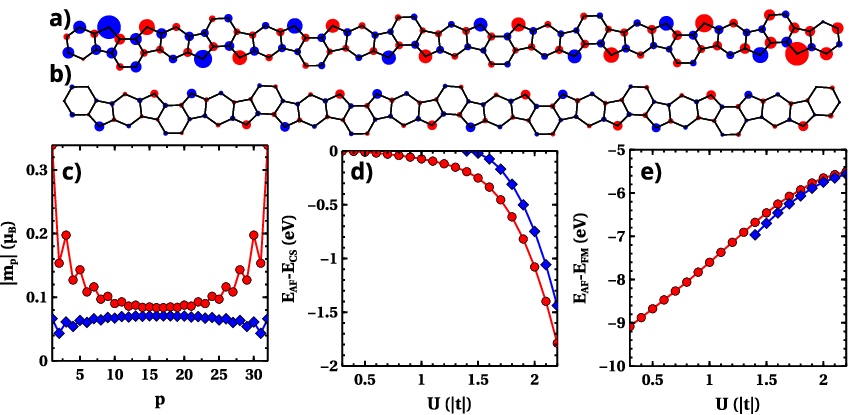}
\caption{Local magnetic moments per site calculated with a mean-field approximation of the Hubbard model for double isomeric Class-II OInIn polymers with terminations a) hosting edge states and b) without edge states, colour stands for the moment sign. c) Magnetic moment at each pentagon vertex $p$ for polymers with both terminations and $P=34$. Calculations were done with $U=1.4|t|; t=-2.7$eV. d)-e) Calculation of the energy difference between the antiferromagnetic solution and the ferromagnetic and non-magnetic solutions as a function of $U$. Circles (rhombus) are for polymers with (without) edge states.}
%\label{Structure}
\label{fig4}
\end{figure*}

As we can see in Figure\ref{fig2}c-d, the resulting tight-binding band structure of the double isomeric polymer presents a direct gap, which has its origin in the alternation of the effective hopping between adjacent unpaired electrons. To ensure numerical stability in the presence of band crossings, along with a careful selection of a unit cell that keeps inversion/mirror symmetry, the Zak phase at half filling was computed using the Wilson loops formalism\cite{resta1999macroscopic,yu2011equivalent} (see supp. mat.). As in the SSH model, the value of ${\cal Z}\in\{0,\pi\}$ was highly dependent on whether the unit cell terminated with the orientation of one of the two isomers from Fig.\ref{fig1}, indicating two different topological regimes.  

\begin{figure}[h!]
 \centering
    \includegraphics[width=0.48\textwidth]{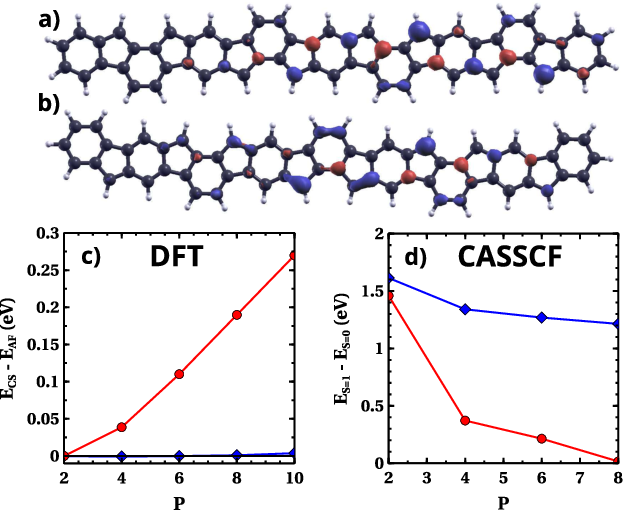}
\caption{a)-b) HOMO isosurface calculated with DFT for the spin-up channel in a molecule with $P=8$ for each termination. Calculations as a function of the number of pentagons $P$ of double isomeric Class-II OInIn polymers of c) the energy difference between the antiferromagnetic and the closed-shell solutions calculated with DFT and d) the first excitation energy calculated with CASSCF. 
Circles (rhombus) are for polymers with (without) edge states.}
%\label{Structure}
\label{fig5}
\end{figure}

An immediate consequence of non-trivial topology in one dimension is the presence of states strongly localized at the chain termini\cite{toyAKLT}. By computing the tight-binding spectrum of finite molecules, I can show that the proposed double isomeric OInIn hosts such kind of edge states depending on the termination (Fig.\ref{fig3}a-c). These topologically protected states, unlike trivial HOMO and LUMO, are separated by an energy splitting ($\delta$) that tends to zero by increasing the number of pentagons (Fig.\ref{fig3}d), which is the consequence of the exponential decay towards the center of the chain (Fig.\ref{fig3}e), also characteristic of topological states\cite{toyAKLT}.

In the following, I consider electronic interactions with the mean-field approximation of the Hubbard model. In carbon-based systems, if $\delta$ is low enough the molecule turns magnetic for a physically feasible value of $U$, displaying local magnetic moments mainly localized at the atoms hosting $\varphi_z(i)$\cite{Yazyev_2010,JFR07}. In Fig.\ref{fig4}a-c, we show the calculated magnetic moments for the OInIn polymer under study, where we can clearly see that, depending on the termination, the distribution of the moments changes dramatically. This is a direct consequence of the two different topological regimes, which also affects magnetism. Since a ribbon with edge states will present a much lower HOMO-LUMO separation $\delta$, it will turn magnetic for a smaller value of $U$ than a molecule without them, presenting magnetic moments mainly localized at the edges. As we can see in Fig.\ref{fig4}d-e for a  $P=16$ system, as soon as the molecule is magnetic, the antiferromagnetic phase is the most stable solution, confirming that the dominant magnetic exchange couplings are hopping-mediated between first-neighbours\cite{jacob2022theory}, as it was predicted by the Class-II classification of the two isomeric constituents\cite{ortiz2023magnetic}.

Finally, I performed quantum chemistry calculations to explore the presence of edge states and magnetism in realistic molecules. OInIn with even $P$ are kekulenic molecules, hence it is possible to draw the double bonds both in an open- and a closed-shell configuration. Actually, it is predicted by theory\cite{JACSN2,C4CP03169E}, and confirmed with atomic force microscopy experiments for one of them\cite{transdimerexp}, that the smallest Class-II OInIn (i.e. indenofluorenes) are closed-shell (Fig.\ref{fig5}c), evolving into an antiferromagnetic open-shell system when $P$ is increased\cite{transchainexp,yamane2018open,fukuda2017theoretical}. This happens because in Class-II OInIn there is just a gain of one Clar sextet for each pair of unpaired electrons, but delocalization stabilizes the antiferromagnetic solution for larger polymers. 

I optimized the geometries of molecules of different lenght with DFT using PBE0/def2-SVP for both the restricted and unrestricted singlet spin multiplicities. As in the model Hamiltonians, the disparities found in DFT depending on the termination can be attributed to the diferent topological regimes, showing a strong localization of the HOMO at the termini of the molecules corresponding to the non-trivial case (Fig.\ref{fig5}a-b). The presence of edge states heavily influences the magnetic behaviour also in realistic structures (Fig.\ref{fig5}c). As a direct consequence of a smaller $\delta$, non-trivial molecules turn magnetic already at $P\geq 4$, whereas trivial molecules need more pentagons for the antiferromagnetic solution to be stabilized, being the closed-shell and open-shell solutions still mostly degenerate even for $P=10$ ($E_{CS}-E_{AF} \sim 3$meV). In all cases, the open-shell $S=0$ multiplicity was more stable than the ferromagnetic high-spin solution.

The spin unrestricted Kohn-Sham wave functions, calculated with DFT, were later used as input for the multireference CASSCF method. I included at least $P$ electrons in $P$ active orbitals for the active space (see supp. mat.), which capture the essential physics in multiradical systems. Molecules with an odd number of pentagons (i.e., $P = 3$ and $5$) hosted an $S=1/2$ ground state with the dominant configuration presenting one single-occupied natural orbital. This is consistent with the Class-II classification, displaying the effective low energy physics of odd-numbered tight-binding chains with one zero mode and an antiferromagnetic ground state.

Even-numbered molecules, on the other hand, may be closed- or open-shell. As it happened with the other methods, I obtained different qualitative results depending on the edge termination. In both cases, the shortest molecules ($P=2$) had a considerable singlet-triplet excitation energy ($\Delta_{TS}=E_{S=1}-E_{S=0}\approx 1.5$eV), where the most relevant configuration in the $S=0$ ground state had closed-shell character. For the non-trivial molecules, $\Delta_{TS}$ decreased exponentially tending to zero as $P$ increased (Fig.\ref{fig5}d), while configurations with orbitals occupied each by just one electron gained weight, and two natural orbitals corresponding to the topologically protected states approached in energy. On the other hand, for the trivial termination, due to the absence of edge states, $\Delta_{TS}$ did not present such a pronounced decrease, which is compatible with a transition to a magnetic regime as $P$ increased with a weaker open-shell character (Fig.\ref{fig5}d).

\section{Conclusions}

In this work, I have computed the low-energy electronic and magnetic properties of an OInIn consisting in the alternating repetition of two different Class-II isomers. Since the dominant coupling in this class corresponds to hopping between adjacent pentagon unpaired electrons, the effective low-energy physics was expected to be that of a simple tight-binding chain. By combining to different isomers, two effective hoppings need to be included, consequently opening a gap in the otherwise gapless non-interacting tight-binding band structure. This gap openning is consistent with the physics of the SSH chain, showing two topological regimes that could be characterized by the Zak phase, depicting topologically protected edge states depending on the edge terminations. 

In the single-particle picture, electronic interactions were first introduced by a mean-field Hubbard model. The inclusion of the on-site Coulomb interaction revealed that the finite fragments are prone to magnetism with an open-shell $S=0$ antiferromagnetic solution as the ground state. The local magnetic moments presented more weight on the pentagons due to the electronic localization of the unpaired electrons wave functions, presenting an exponential decay towards the center of the molecule in the non-trivial case as a consequence of the topological edge states.

More realistic molecules were computed by optimizing their geometries with DFT. In this approach I also obtained strongly localized edge states, supporting the findings in the non-interacting approximation. The magnetic behaviour was very different depending on the edge termination. For the non-trivial case, the molecules became clearly magnetic for $P\geq 4$ because of the smaller energy separation $\delta$ of the edge states, in comparison with molecules of the trivial termination with a higher $\delta$ that were non-magnetic until $P=8$. In the same line, multireference CASSCF calculations also reflected the presence or not of edge states, where the triplet-singlet energy separation ($\Delta_{TS}$) converged differently as a function of the number of pentagons.   

\section{Acknowledgements}

I thank Manuel Melle Franco for fruitful discussions.
This work was supported financially within the scope of the
project CICECO-Aveiro Institute of Materials, UIDB/50011/
2020 (DOI 10.54499/UIDB/50011/2020), UIDP/50011/
2020 (DOI 10.54499/UIDP/50011/2020), and LA/P/0006/
2020 (DOI 10.54499/LA/P/0006/2020), financed by national
funds through the FCT/MCTES (PIDDAC). This work has received funding from the
European Union’s Horizon 2020 research and innovation
program, under grant agreement No. 101046231,
and from the Foundation for Science and Technology (FCT)
under grant agreement M-ERA-NET3/0006/2021 through the
M-ERA.NET 2021 call.

\bibliographystyle{apsrev-title}
\bibliography{references.bib}

\end{document}